\documentclass[11pt]{cernrep}
\usepackage{epsfig}
\usepackage{here}
\usepackage{amsmath}
\usepackage{wasysym}
\def\CC{\hbox{\it l\hskip -5.5pt C\/}}
\begin{document}
 \title{FIELDS ON PARACOMPACT MANIFOLD AND ANOMALIES}
\author{Pierre Grang\'e $^{(a)}$  and Ernst Werner $^{(b)}$}
\institute{$(a)$ Laboratoire de Physique Math\'ematique UMR-CNRS $5825$,\\
 Universit\'e de Montpellier II,$34060$ Montpellier-Cedex, France\\
$(b)$ Institut f\"{u}r Theoretische Physik, Universit\"{a}t Regensburg,\\
D-93040 Regensburg, Germany}
\maketitle
\begin{abstract}
In Continuum Light Cone Quantization (CLCQ) the treatment  of scalar fields
 as operator valued distributions and properties of the accompanying test functions 
are recalled. Due to the paracompactness property of the Euclidean manifold
 these  test functions appear as decomposition of unity. The approach is extended to QED 
 Dirac fields in a gauge invariant way. With such test functions the usual triangle
anomalies are calculated in a simple and transparent way. 
\end{abstract}
\section{INTRODUCTION}
Over the years the interest in Light Cone (LC) formulation of field theories keeps growing mainly because
of the varieties of physical processes amenable to direct evaluation, as reported in this volume. An
important issue still under debate is the treatment of LC induced infrared (IR) divergencies. Compactification
in one LC direction, say $x^{-}=t-x^{3}$, with appropriate boundary conditions, permits an {\it{ad hoc}} elimination
of the problematic zero mode  of the field operator. But it is well recognized by now that it is precisely this 
zero mode which carries the important non-perturbative informations which, in the equal time formalism, are
present in the existence of a non-trivial vacuum. For many purposes, in particular to study critical properties 
of a given field theory, a non-compact formulation is necessary. It uses the notion of fields as operator valued
distributions (OPVD) as developped in [1]. These studies focussed on $\Phi^{4}$ scalar field theory in $1+1$ dimension.
Here we want to extend this approach to gauge theories. Due to the paracompactness property of an Euclidean manifold
we show that the OPVD formulation permits a simple and transparent evaluation of the QED triangle anomalies.
\section{FIELDS AS  OPVD}
The Klein-Gordon (KG) equation for the free scalar field in D-dimension , $(\Box_x + m^2)\varphi(x) = 0$, writes, after a Fourier 
transform, $(p^2-m^2)\tilde{\varphi}(p)=0$. The solution is a distribution $\tilde{\varphi}(p)=\delta^{(D)}(p^2-m^2)
\chi(p)$, with $\chi(p)$ arbitrary. The solution of the KG-eqution is therefore also a distribution, {\it{ie}} an 
OPVD, which defines a functional with respect to a test function $\rho(x)$, which is $C^{\infty}$ with compact
support,
\begin{equation}
\Phi(\rho) \equiv <\varphi,\rho> = \int d^{(D)}y\varphi(y)\rho(y).
\end{equation}
$\Phi(\rho)$ is a $C$-number with the possible interpretation of a more general functional $\Phi(x,\rho)$ evaluated at $x=0$.
Indeed the translated functional is a well defined object [2] such that 
\begin{equation}
T_{x}\Phi(\rho) = <T_{x}\varphi,\rho> = <\varphi,T_{-x}\rho> = \int d^{(D)}y \varphi(y) \rho(x-y)
\end{equation}
Now the test function $\rho(x-y)$ has a well defined Fourier decomposition
\begin{equation}
\rho(x-y) = \int{d^{(D)}q \over (2 \pi)^{D}}\exp^{iq(x-y)}f(q)
\end{equation}
It follows that
\begin{equation}
T_{x}\Phi(\rho) = \int{d^{^(D)}p \over (2 \pi)^{D}} e^{-ipx}\delta(p^2-m^2)\chi(p)f(p).
\end{equation}
Due to the properties of $\rho$ $, T_{x}\Phi(\rho)$ obeys the KG equation and is taken as the physical field with quantized
form
\begin{equation}
\varphi_{1}(x) = \int{d^{(D-1)}p \over (2 \pi)^{(D-1)}}{1 \over \sqrt{2 \omega_{p}}}[a^{+}_{p} e^{ipx} + a_{p} e^{-ipx}]f(p,\omega_{p}).
\end{equation}
$f(p,\omega_p)$ acts as regulator with very specific properties \footnote{f(p) is also $C^{\infty}$ with fast decrease
in the sense of L. Schwartz [2].}[4]. This expression for $\varphi_{1}(x)$ is particurlarly useful on the LC because the Haag 
serie can be used and is well defined in terms of products of $\varphi_{1}(x_i)$.
\section{PARACOMPACT MANIFOLD: TEST FUNCTIONS AS DECOMPOSITION OF UNITY}
Consider a topological space $\mathcal{M}$. An open covering [3] of $\mathcal{M}$ is a family of open subspaces $\Omega_i$, $i \in I$, with the
property $\mathcal{M} = \bigcup_{i \in I} \Omega_i$. Paracompactness is the property
that for each $\Omega_i$ there exists a $C^{\infty}$ function $\beta_{i}(x)$
such that $\beta_{i}(x) = 1$ if $x \in \omega_i \subset  \Omega_i$, $0 < \beta_{i}(x) < 1$ in the boundary region
 $ \omega_{i} \subset \mathcal{B}_i \subset \Omega_i$, and $\beta_{i}(x) = 0$ {\it{outside}} $\Omega_i$. For all $x \in \mathcal{M}$ there is only a finite number of $\beta_{j}(x) \not= 0$. Let $\alpha_{j} = {\beta_{j} \over \Sigma_{j}
\beta_j}$. Now $\Sigma_{j} \beta_j$ is always non zero and $\Sigma_{j} \alpha_j = 1$. $\{\alpha_j\}$ is therefore a
 decomposition of unity on $\mathcal{M}$ \footnote{An explicit construction involves the characteristic function $\chi_{\omega_{j}}(x) = 1(0)$ if $x \in(\notin)
~ \omega_{j}$ and Schwartz's test function $\rho_{\epsilon}(x)$ in the ball $\mathcal{B}(\epsilon)$ ,
 $\beta_{j}(x) = \int \chi_{\omega_{j}}(t) \rho_{\epsilon}(x-t) dt$.}. The important theorem is that: "An Euclidean manifold is
 paracompact" [3]. We shall therefore work in Euclidean metric. Then $f(p)$ is $1$ except in the boundary region where it is
$C^{\infty}$ and goes to zero with all its derivatives.
\section{QED: CONSTRUCTION AND GAUGE TRANSFORMATION OF THE OPVD FERMIONIC FIELD}
Let $\psi(x)$ be the Dirac massive free field, then $(i\not{\!\!\!\partial} - m)\psi(x)=0 \Longrightarrow \Psi(x)\equiv
<T_{x}\psi,\rho>=\int d^{(D)}y\psi(y)\rho(y-x)$. For QED the fermionic field obeys $(i\not{\!\!\partial}-\not{\!\!\!A}- m)\psi(x)=0$,
and it is clear that the translation in $\Psi(x)$ must be done in a gauge invariant way, that is
\begin{equation}
\Psi_{\gamma}(x) = \int d^{(D)}y \rho(y-x) \exp [i e \int^{y}_{x} dz^{\mu} A_{\mu}(z)] \psi(y).
\end{equation}
In a gauge transformation $A_{\mu}(x) \rightarrow A_{\mu}(x) +{1\over e} \partial_{\mu} \Lambda(x)$,$ \psi(y) \rightarrow
 e^{i\Lambda(y)}\psi(y)$ and then $\Psi_{\gamma}(x) \rightarrow e^{i\Lambda(x)}\Psi_{\gamma}(x)$. Due to the presence of
the gauge phase factor in (6) $\Psi_{\gamma}(x)$ is path dependant. Let $\gamma(s)$ be a parametrization of the path from
 $x$ to $y$, $\gamma(0)=x, \gamma(1)=y$. Then
\begin{eqnarray}
 \int^{y}_{x} dz^{\mu} A_{\mu}(z) & = & \int_{0}^{1} ds \dot{\gamma}^{\mu}(s) A_{\mu}(\gamma(s)) =  \int d^{(D)}z [\int_{0}^{1} ds \dot{\gamma}^{\mu}(s) \delta^{(D)}(\gamma(s)-z)] A_{\mu}(z) \nonumber \\
                             & \equiv & \int d^{(D)}z~\CC^{\mu}(\gamma ; x,y,z) A_{\mu}(z ) 
\end{eqnarray}
It is easy to see that $\CC^{\mu}(\gamma ; x,y,z)$ obeys the differential equation $\partial_{z}^{\mu}
\CC_{\mu}(\gamma ; x,y,z)=\delta(x-z)-\delta(y-z)$, the solution of which is known only after a choice of
path and boundary condition on z \footnote{Solution of the form [4] $\CC^{\mu}(\gamma ; x,y,z)=\partial^{\mu}
c(\gamma;x,y,z)$ are excluded, for then $\int d^{(D)}z(\partial_{z}^{\mu}c)A_{\mu}=-\int d^{(D)}z c (\partial_{z}^{\mu}
A_{\mu})$ which would be zero in the Lorentz gauge.}. With $y-x=\epsilon$, the OPVD Dirac field is now
$\Psi_{\gamma}(x) = \int d^{(D)}\epsilon \rho(\epsilon) \exp [i e \int^{x+\epsilon}_{x} dz^{\mu} A_{\mu}(z)]
\psi(x+\epsilon)$. One expects that if the extent of the ball $\mathcal{B}(\epsilon)$, support of $\rho(\epsilon)$,
is "small" the straight path is the good choice. This is corroborated when evaluating the change $\Delta \Psi_{\gamma}(x)=
\Psi_{\gamma+\delta \gamma}(x)-\Psi_{\gamma}(x)$ for a change $\delta \gamma$ of the path $\gamma$. Indeed 
 $\Delta \Psi_{\gamma}(x) \propto ( \Delta \gamma^{\nu}(s)\Delta \dot{\gamma}^{\mu}(s)-  \Delta \gamma^{\mu}(s)\Delta 
\dot{\gamma}^{\nu}(s))F_{\nu,\mu}(\gamma(s)) \Psi_{\gamma}(x)$, which is zero for a straight path $\Delta \gamma(s)=f(s)(y-x), f(0)=0, f(1)=1$.
\section{QED ANOMALIES}
We consider the usual QED triangle diagrams with Ryder's convention [5] and Euclidean metric. Let $I^{1}_{\kappa,\lambda,
\mu}$ and $I^{2}_{\lambda,\kappa,\mu}$ be the direct and exchange contributions respectively. The direct axial current
 contribution writes, after performing the traces on $\gamma-$matrices
\begin{equation}
(p_{1}+p_{2})^{\mu} I^{1}_{\kappa,\lambda,\mu} = 4e^{2}\epsilon_{\sigma,\lambda,\delta,\kappa}\int {d^{4}k \over
 (2\pi)^4}\Big{[}{p_{2}^{\sigma} k^{\delta} \over (k+p_2)^2 k^2 }- {k^{\delta} p_{1}^{\sigma} \over k^2 (k-p_1)^2 } \Big{]}
f(k^2)f((k+p_2)^2)f((k-p_1)^2),
\end{equation}
where the $f$'s factors come from the test fuctions present in the fermionic propagators to lowest order in $e$ and 
$\epsilon_{\sigma,\lambda,\delta,\kappa}$ is the usual antisymmetric tensor.
The exchange axial contribution is obtained with the changes $(\kappa \leftrightarrow \lambda), (p1 \leftrightarrow
p2)$. Due to the $f$'s the integrals are finite: one may change $k$ to $k-p_{1}$ in the first integral and
$k$ to $k+p_{2}$ in the second. Regrouping terms the total axial contribution is now
\begin{eqnarray}
(p_{1}+p_{2})^{\mu}( I^{1}_{\kappa,\lambda,\mu}+I^{2}_{\lambda,\kappa,\mu})&=&4e^{2}\epsilon_{\sigma,\lambda,\delta,\kappa}
\int {d^{4}k \over (2\pi)^4}{k^{\delta} \over k^2}\Big{\{}{ p_{1}^{\sigma} \over  (k-p_1)^2 }\Big{[}f((k-p_{1}-p_{2})^2)
-f((k+p_{2})^2)\Big{]} \nonumber\\
f((k-p_{1})^2)&-&{ p_{2}^{\sigma} \over (k+p_2)^2 }\Big{[}f((k+p_{1}+p_{2})^2)-f((k-p_{1})^2)\Big{]}f((k+p_{2})^2)\Big{\}}f(k^2).
\end{eqnarray}
It is seen that if $f=1$ everywhere the axial contribution would be zero, but the variable change in this case is not 
legitimate for the integrals are linearly divergent. However $f=1$ {\it{almost}} everywhere except in the vicinity of
the boundary of its support. Its generic shape in the $k_\delta$ direction (in dimensionless units)
is shown in FIG.1. Clearly the situation of interest is the large $\Lambda$ limit
\vspace*{-3cm}
%%%%%%%%%%%%%%%%%%%%%%%%%%%%%%%%%%%%%%%%%%%%%%%%%%%%%%%%%%%%%%%%%%
\vspace{0.9cm}\hspace*{1.5cm}\psfig{file=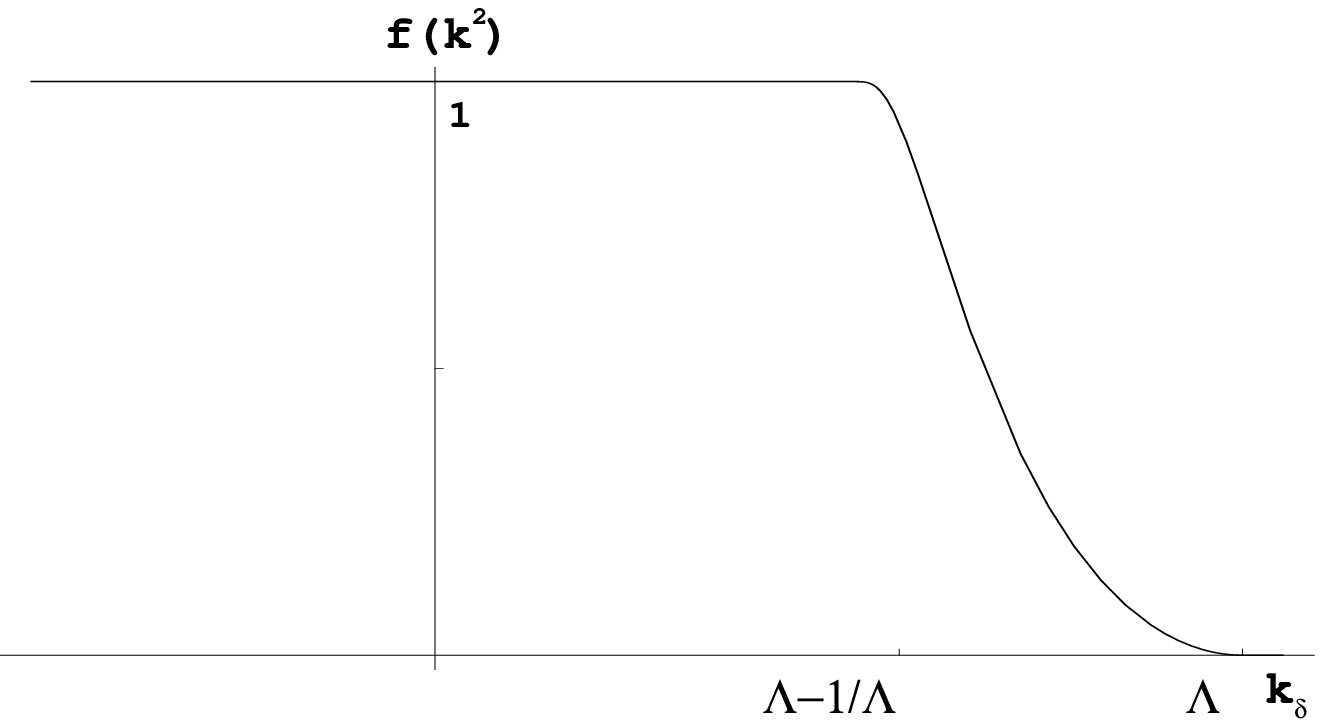,height=5cm,width=6cm}\hspace*{1.7cm}\psfig{file=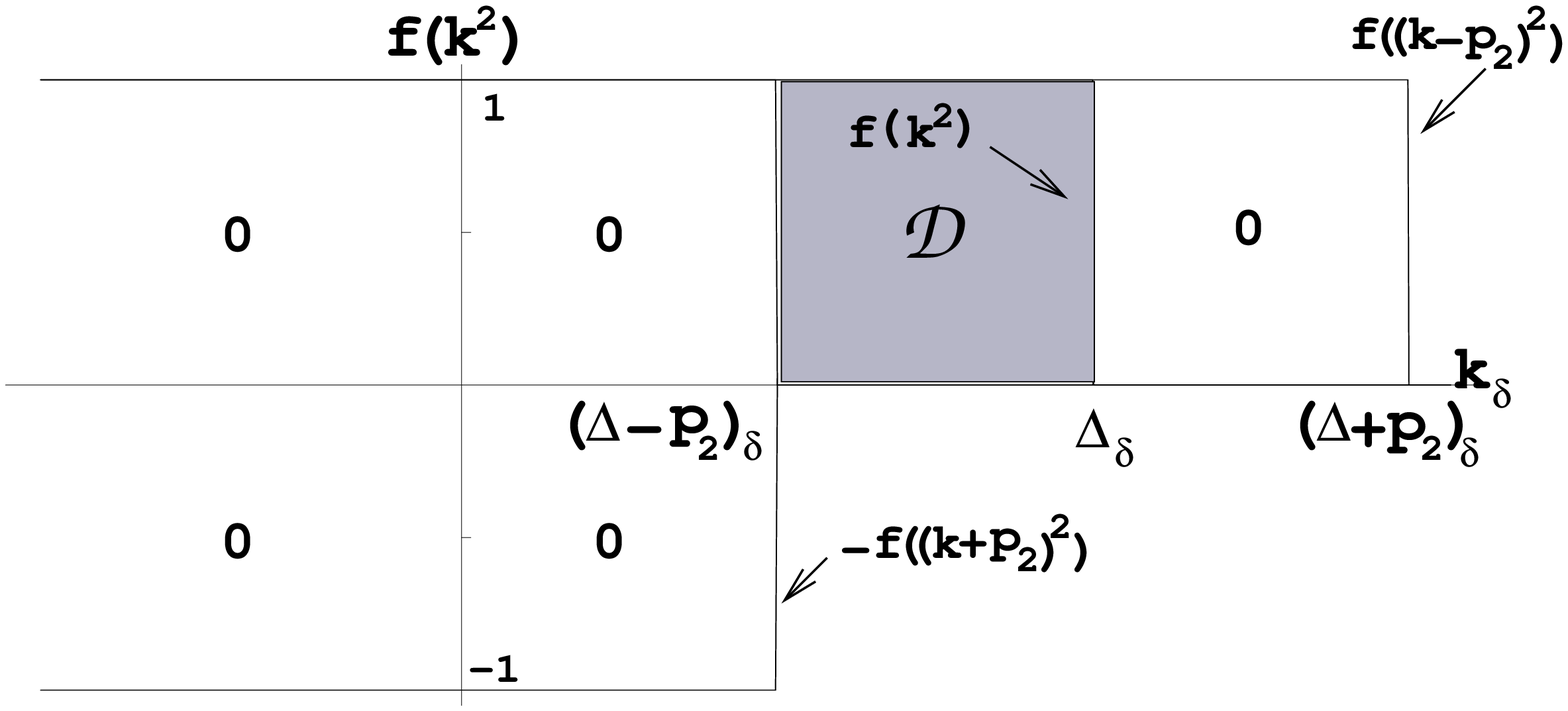,height=5cm,width=6cm}\\
\par
\vspace*{1cm}\hspace*{0.0cm}{FIG. 1 Generic shape of $f(k)$ as a function of $k$.}\hspace*{1cm}{FIG. 2 Domain $\mathcal{D}$ where $\Delta f \neq 0$}\\
%%%%%%%%%%%%%%%%%%%%%%%%%%%%%%%%%%%%%%%
\par
and we can look at cases where $p_{1},p_{2} \ll \Lambda$
and all $f$'s shrink to step functions.Then Eq.(9) reduces to
\begin{eqnarray}
(p_{1}+p_{2})^{\mu}( I^{1}_{\kappa,\lambda,\mu}+I^{2}_{\lambda,\kappa,\mu})=4e^{2}\epsilon_{\sigma,\lambda,\delta,\kappa}
\int {d^{4}k \over (2\pi)^4}{k^{\delta} \over k^4}f^2(k^2)\Big{\{} p_{1}^{\sigma}\Big{[}f((k-p_{2})^2)-f((k+p_{2})^2)
\Big{]}\nonumber\\
-p_{2}^{\sigma}\Big{[}f((k+p_{1})^2)-f((k-p_{1})^2)\Big{]}\Big{\}}.
\end{eqnarray}
Consider the quantity $\Delta f= f^2(k^2)\Big{[}f((k-p_{2})^2)-f((k+p_{2})^2)\Big{]}$ in the direction of $k_\delta$. The situation is depicted
in FIG.2.
$\Delta f $ is different from zero in the shaded area $\mathcal{D}$ of amplitude $(p_2)_\delta$, $\forall \Lambda$. Hence
\begin{equation}
\int_{(\Lambda-p_2)_\delta}^{\Lambda_\delta}{d^{4}k \over (2\pi)^4}{k^{\delta} \over k^4}={2\pi^2 \over(2\pi)^4}
\int_{(\Lambda-p_2)_\delta}^{\Lambda_\delta}
dk{k_\delta \over k}={1 \over 8\pi^2} \int_{(\Lambda-p_2)_\delta}^{\Lambda_\delta}dk_\delta={(p_2)_\delta \over 8\pi^2},
\end{equation}
and we have the result $(p_{1}+p_{2})^{\mu}( I^{1}_{\kappa,\lambda,\mu}+I^{2}_{\lambda,\kappa,\mu})={e^{2} \over 2\pi^2} \epsilon_{\sigma,\lambda,\delta,\kappa}
\Big{[} p_1^\sigma p_2^\delta + p_2^\sigma p_1^\delta\Big{]}=0$,
because of the antisymmetry of $\epsilon_{\sigma,\lambda,\delta,\kappa}$. The axial current is therefore conserved.
Consider now the vector current. After tracing over the $\gamma-$matrices the potentially divergent contribution is
\begin{eqnarray}
p_1^\kappa(I^{1}_{\kappa,\lambda,\mu}+I^{2}_{\lambda,\kappa,\mu})=-4e^{2}\epsilon_{\lambda,\kappa,\mu,\alpha}p_1^\alpha
\int {d^{4}k \over (2\pi)^4}{k^{\alpha} \over k^4}f(k^2)\Big{[}f((k-p_1)^2)f((k+p_2)^2) \nonumber\\
-f((k+p_1)^2f((k-p_2)^2)\Big{]}.
\end{eqnarray}
Denote $\Delta f$ the test function factor and let $e^\nu (\psi_k,\varphi_k,\theta_k)={k^\nu \over k}=\{\sin(\psi_k) \sin(\varphi_k) \cos(\theta_k),etc\}$
Performing the analysis of $\Delta f$ in terms of step functions gives, using $\cos{\theta}_{k p_i}= p_i^\nu e_\nu / p_i$,
$\int dk \Delta f=2(p_1\cos\vartheta_{k p_1}-p_2\cos\vartheta_{k p_2})=2(p_1-p_2)^\nu e_\nu(\psi_k,\varphi_k,\theta_k)$.
The integral over $d\Omega_k$ is now straightforward with the result $p_1^\kappa(I^{1}_{\kappa,\lambda,\mu}+I^{2}_{\lambda,\kappa,\mu})={e^{2}\over 4\pi^2} \epsilon_{\kappa,\lambda,\mu,\alpha} p_1^\kappa p_2^\alpha.$
The vector current (charge) conservation is therefore restored with the correction 
$\delta I_{\kappa,\lambda,\mu}={e^2 \over 4\pi^2} \epsilon_{\kappa,\lambda,\mu,\alpha} (p_1-p_2)^\alpha$, resulting in the standard axial anomaly $(p_{1}+p_{2})^{\mu}( I^{1}_{\kappa,\lambda,\mu}+I^{2}_{\lambda,\kappa,\mu}+\delta I_{\kappa,\lambda,\mu})={e^{2} \over 2 \pi^2}\epsilon_{\kappa,\lambda,\mu,\alpha} p_2^\mu p_1^\alpha$.
\section{CONCLUSIONS}
Treating scalar fields as OPVD gives a consistent LCQ in the continuum which permits the study of critical
 properties. It is achieved because IR-induced divergences are handled by the test function present in the 
regularized field which, in the limit $k^+ \rightarrow 0$, goes to zero faster than any inverse power of $k^+$.
An essential feature is also the possible use of the Haag serie, for its construction is well defined in terms of
the regularized scalar field. In going to gauge theories the definition of the regularized
fermionic field from its OPVD counterpart faces the problem of gauge invariance. Taking into account the necessity 
that the original OPVD fermionic field must be translated in a gauge invariant manner leads to a regularization 
scheme which does not suffer the general illness of a straight momentum cut-off. It is examplified in the field 
equation, $\partial_\mu F^{\mu,\nu}(x)=j^{\nu}(x)= \bar{\Psi}_{\gamma}(x)\gamma^\nu\Psi_{\gamma}(x)$, since by
construction the regularized fermionic field renders the current $j^{\nu}(x)$ gauge invariant. The important property 
of paracompactness of the Euclidean manifold permits using test functions which are decomposition of unity. They lead to a transparent analysis of the QED anomalies, in complete agreement with the standard results. This is a strong 
incitation to pursue further the investigations on the merits and possible illnesses of this regularisation scheme in
 the context of the LC formalism of field theories.
\section*{ACKNOWLEDGEMENTS}
Simon Dalley is gratefully acknowleged  for organizing  this stimulating meeting. E. Werner thanks A. Neveu for his 
interest in the LC formalism and financial support of visits to Montpellier.

\end{document}